\documentstyle[twocolumn,prl,aps,epsfig]{revtex}

\begin{document}
\draft
\twocolumn[\hsize\textwidth\columnwidth\hsize\csname @twocolumnfalse\endcsname
%
%
%

\title{Elementary excitations of the symmetric spin-orbital model: The
XY limit}

\author{Fr\'ed\'eric Mila $^1$, Beat Frischmuth $^2$, Andreas Deppeler $^2$ and 
Matthias Troyer $^3$}
\address{$^1$ Laboratoire de Physique Quantique, Universit\'e Paul Sabatier, 118 
route de Narbonne, F-31062 Toulouse Cedex\\
$^2$ Institute of Theoretical Physics, ETH H\"onggerberg, 
CH-8093 Z\"urich, Switzerland \\
$^3$ Institute for Solid State Physics, University of Tokyo, Roppongi 7-22-1,
Tokyo 106 }

\date{\today}
\maketitle

\begin{abstract}
  The elementary excitations of the 1D, symmetric, spin-orbital model are
  investigated by studying two anisotropic versions of the model, 
  the pure XY and the dimerized XXZ case, with analytical and numerical 
  methods. While they preserve the symmetry between spin and orbital degrees of
  freedom, these models allow for a simple and transparent picture of the
  low--lying excitations: In the pure XY case, a phase
  separation takes place between two phases with free--fermion like, 
  gapless excitations, 
  while in the dimerized case, the low-energy effective
  Hamiltonian reduces to the 1D Ising model with gapped excitations.
  In both cases, all the elementary excitations involve simultaneous
  flips of the spin and orbital degrees of freedom, a clear indication
  of the breakdown of the traditional mean-field theory.
\end{abstract}

\vskip2pc]
\narrowtext

The impact of orbital degeneracy on the low--energy properties of
Mott--Hubbard insulators is currently attracting a lot of attention
following the progress in synthetizing and studying materials with
these characteristics \cite{bao}. It was already pointed out a long time ago by
Kugel and Khomskii that such systems should have low--lying orbital
excitations in addition to spin excitations \cite{kugel}. More recently, it has
been suggested that the interplay between both degrees of freedom can
have more dramatic consequences. For instance, under suitable
conditions the orbital degeneracy can enhance quantum fluctuations in
the spin degrees of freedom and lead to gapped spin excitations even
in the 3D case\cite{feiner}. Another interesting situation is the SU(4) symmetric
case\cite{arovas,zhang} where the system cannot choose locally between 
the configurations
(spin singlet $\times$ orbital triplet) and (spin triplet $\times$
orbital singlet). Then the mean-field approach that decouples spin and
orbital degrees of freedom on each bond cannot be a good starting
point in that case since it violates basic SU(4) relationships between
correlation functions on a given bond, as emphasized in
Ref.\cite{frischmuth}. As a consequence, the traditional picture of
relatively independent spin and orbital excitations must be abandoned.
A clear picture of the low-lying excitations in such a case is still
lacking though.

In this Letter, we concentrate on the symmetric case. The
basic model is the SU(4) symmetric Hamiltonian given by
\begin{equation}\label{ham1}
H=J\sum_i \left(2\vec{S}_i\cdot\vec{S}_{i+1}+\frac{1}{2}\right)
          \left(2\vec{\tau}_i\cdot\vec{\tau}_{i+1}+\frac{1}{2}\right)
\end{equation}
where $\vec S_i$ and $\vec \tau_i$ are spin-1/2 operators corresponding to spin and
orbital degrees of freedom respectively.
This model has already been studied rather extensively by several
methods \cite{arovas,zhang,frischmuth,sutherland,affleck,japan}. 
In particular, it is known from the 
Bethe ansatz solution that there
are three branches of low-energy excitations\cite{sutherland}. The
physical interpretation of these branches is not straightforward
though. The essential complexity comes from the large local
degeneracy: For a single bond, the groundstate is six-fold degenerate
(spin singlet $\times$ any of the three orbital triplets or any of the
spin triplets $\times$ orbital singlet). It is thus interesting to study the
XXZ version of the model defined by
\begin{eqnarray}\label{ham2}
H & = & \sum_i J_i \left(2(S_i^x S_{i+1}^x + S_i^y S_{i+1}^y + 
\lambda S_i^z S_{i+1}^z ) + \frac{\lambda}{2}\right)\nonumber \\
& & \times \left(2(\tau_i^x \tau_{i+1}^x + \tau_i^y \tau_{i+1}^y + 
\lambda \tau_i^z \tau_{i+1}^z )+ \frac{\lambda}{2}\right)
\end{eqnarray}
In that case, the degeneracy is lifted within the triplet sector as
soon as $\lambda<1$, and the groundstate of a given bond is only
two-fold degenerate (spin singlet $\times$ orbital triplet with
$\tau_{tot}^z=0$ or spin triplet with $S_{tot}^z=0$ $\times$ orbital
singlet).  The essential ingredient, namely the symmetry between spin
and orbital degrees of freedom, is preserved, but the Hilbert space of
the low-lying sector is considerably reduced.

In the following, we will concentrate on two versions of this model
for which a transparent picture of the low-lying excitations can be
obtained:

{\bf The pure XY model:} It corresponds to the previous Hamiltonian
[Eq. (\ref{ham2})] with $\lambda=0$ and $J_i=J$ for all bonds, which
can be written more compactly as
\begin{equation}\label{ham3}
H=J\sum_i \left(S_i^+ S_{i+1}^- + S_i^-S_{i+1}^+ \right)
\left(\tau_i^+ \tau_{i+1}^- + \tau_i^-\tau_{i+1}^+ \right)
\end{equation}
or in expanded form
\begin{eqnarray}\label{ham4}
H&=&J\sum_i \left(S_i^+\tau_i^+S_{i+1}^-\tau_{i+1}^- + 
S_i^-\tau_i^-S_{i+1}^+\tau_{i+1}^+\right) \nonumber \\
& &+ J\sum_i \left(S_i^+\tau_i^-S_{i+1}^-\tau_{i+1}^+ + 
S_i^-\tau_i^+S_{i+1}^+\tau_{i+1}^-\right).
\end{eqnarray}

Analyzing this hamiltonian in the product basis $\otimes_i |
\eta\rangle_i$, where $ | \eta\rangle_i = |
S_i^z=\pm\frac{1}{2},\,\tau_i^z=\pm\frac{1}{2}\rangle$ and denoting
\begin{eqnarray}
|a\pm \rangle_i&=& |S_i^z=\pm 1/2,\,\tau_i^z=\pm 1/2\rangle,\nonumber \\
|b\pm \rangle_i&=& |S_i^z=\pm 1/2,\,\tau_i^z=\mp 1/2\rangle, 
\end{eqnarray}
one can easily show that all the matrix elements of $H$ between 
states $\{|a\pm\rangle_i\}$ and $\{|b\pm\rangle_j\}$ vanish.
As a consequence, the eigenstates can be classified according to the 
sequence of domains of {\it  phase A} (with parallel orbital and spin states) 
and {\it phase B} (with antiparallel states) that include only 
states 
of type $|a\pm\rangle_i$ and $|b\pm\rangle_i$ respectively, and 
Eq. (\ref{ham4}) can be written as
\begin{eqnarray}\label{ham5}
H&=&2J\sum_i \frac{1}{2}(\alpha_i^+\alpha_{i+1}^-+\alpha_i^-\alpha_{i+1}^+)
\nonumber \\
& &+ 2J\sum_i \frac{1}{2}(\beta_i^+\beta_{i+1}^-+\beta_i^-\beta_{i+1}^+),
\end{eqnarray}
where $\alpha_i^\pm=S_i^\pm\tau_i^\pm$
($\beta_i^\pm=S_i^\pm\tau_i^\mp$) are the raising and lowering
spin-1/2 operators corresponding to $|a\pm\rangle_i$ ($|b\pm\rangle_i$).
Now the Hamiltonian within a phase can be mapped onto spinless fermions with a
Jordan--Wigner transformation, and the groundstate energy of a domain of length
$L$ is given by  
\begin{equation}\label{finitechain}
E(L)=-2J\cos\left(\frac{(L/2+1)\pi}{2(L+1)}\right) 
\frac{\sin\left(\frac{(L/2)\pi}{2(L+1)}\right)}
{\sin\left(\frac{\pi}{2(L+1)}\right)}
\end{equation}
It is easy to check that $E(L)<E(L_1)+...+E(L_n)$, where $L_1,...,L_n$ are
integers such that $L_1+...+L_n=L$. So the model of Eq. (\ref{ham5}) undergoes a
phase separation, and the groundstate is two-fold degenerate with a single domain
of phase A or
B.

\begin{figure}[t]
  \begin{center}
  \epsfxsize=85mm
  \epsffile{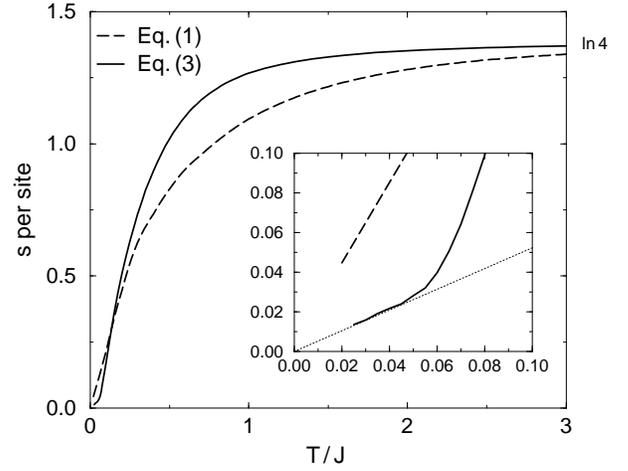}
  \end{center}
  \caption[]{Temperature dependence of the entropy of the SU(4) symmetric 
    spin orbital model of Eq.~(\ref{ham1}) (dashed line, taken from 
    Ref.\cite{frischmuth})
    and of the pure XY spin-orbital model of
    Eq.~(\ref{ham3}) (solid line) for chains of length $L=200$ with periodic
    boundary conditions in two different temperature scales. For
    comparison the analytical result for the entropy of the
    XY-model with coupling $2J$ and infinite length is also shown (dotted
    line).}
  \label{entropy}
\end{figure}

\begin{figure}[t]
  \begin{center}
  \epsfxsize=85mm
  \epsffile{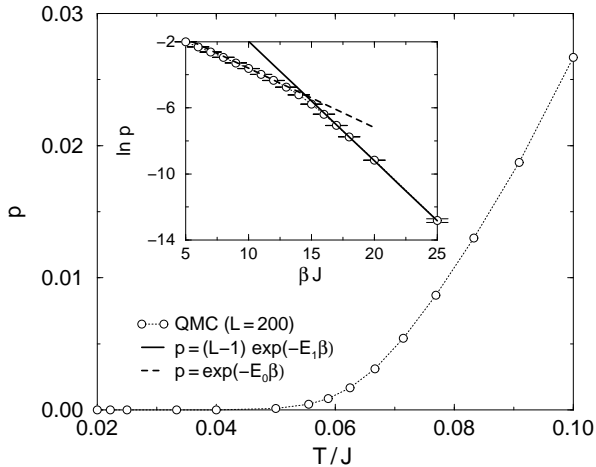}
  \end{center}
  \caption[]{Temperature dependence of the average density $p$ of domain walls  
    in the pure XY spin-orbital model. The error
    bars are smaller than the symbols. The inset shows $\ln p$ as
    function of the inverse temperature $\beta J$ as well as fits
    of the form $p=\exp(-E_0/T)$ and $p=(L-1) \exp(-E_1/T)$ in the 
    intermediate
    and very low temperature ranges (for
    details see text).}
  \label{phase_nr}
\end{figure}

The phase separation has some drastic consequences for the
thermodynamics, and the low-temperature properties of the pure XY spin-orbital
model turn out to be significantly different from those of
the SU(4) symmetric model [Eq.
(\ref{ham1})]. In Fig.~\ref{entropy} we show the
entropies $s$ per site as a function of temperature for both models,
the SU(4) symmetric model [Eq. (\ref{ham1})] and the pure XY-case [Eq.
(\ref{ham3})].  They have been calculated numerically for chains of
length $L=200$ with periodic boundary conditions, using the continuous
time quantum Monte Carlo loop algorithm \cite{evertz,beard}. At very
low temperature, both entropies show a linear behavior (see inset of
Fig.~\ref{entropy}), indicating the presence of gapless excitations. 
But the slope
for the SU(4) is much larger than that in the pure XY spin-orbital
case. In fact, in the first case the slope is three times bigger than
that of single SU(2) antiferromagnetic Heisenberg chain \cite{frischmuth},
while the slope of the entropy at low $T$ in the pure XY spin-orbital
model is equal to that of the XY model with coupling $2J$, 
as expected from Eq.~(\ref{ham5}). A
further difference is visible as the temperature is increased. The
entropy of the SU(4) symmetric spin-orbital model remains
approximately linear also in the intermediate temperature range (up to
$T\approx 0.2 J$).  The entropy of the pure XY spin-orbital model, on
the other hand, coincides with the linear behavior of the XY-model
only up to a crossover temperature $T^*\approx0.05 J$.
Above $T^*$ a sharp increase of the entropy takes place (see inset of
Fig.~\ref{entropy}).  This increase comes from the additional entropy
contribution of the domain walls which are more and more frequent with
increasing temperature. This can be seen in Fig.~\ref{phase_nr} where
the average density  of domain walls $p$ 
(average number of domain walls per site) is depicted as a function of
temperature. Up to $T^*$, the spin-orbital model is in one of the
phases A or B and $p=0$ within the statistical errors of the Monte
Carlo simulation. Above $T^*$ the number of phase sectors increases
very sharply with increasing $T$.

This behaviour can be easily understood in terms of the following,
approximate free energy per site:
\begin{equation}\label{free}
f(T,p) = p E_{DW} - T s_{DW}(T,p) - T  s_D(T)
\end{equation}
The various quantities entering this expression are: i) $p$, the concentration 
of domain walls, to be determined 
by minimizing the free energy; ii)
$E_{DW}$, the energy of a domain wall. This is the energy required to split a
finite chain of length $L$ into two chains of length $L-L_1$ and $L_1$, i.e.
$E(L-L_1)+E(L_1)-E(L)$, where $E(L)$ is defined in Eq. (\ref{finitechain}).
$E_{DW}$ is {\it a priori} a function of $L$ and
$L_1$. It turns out that, for large enough $L$, $E_{DW}\simeq 0.36 J$ regardless 
of $L$ and $L_1$ except for $L_1<L_0\simeq 20$. Since
$L_0$ does not depend on $L$, this difference will play no role in the
thermodynamic limit at low temperatures, and one can safely assume $E_{DW}=0.36 J$; 
iii) $s_{DW}(T,p)$, the entropy of the domain walls. For small $p$, it is given
by $s_{DW}(T,p)=-p\ln p$;
iv) $s_{D}(T)$, the entropy contribution of the domains. When $p$ is small,
finite-size effects are negligible, and $s_D(T)$ is equal to the entropy of the
XY model with coupling $2J$, i.e. $(\pi/6)(T/J)$ at low temperature.

As for the 1D--Ising model, minimizing with respect to the density of domain
walls $p$ leads to
$p=\exp(-E_{DW}/T)$ and $s_{DW}(T)=(E_{DW}/T)\exp(-E_{DW}/T)$. The total 
entropy, which is 
the sum of $s_{DW}(T)$ and $s_D(T)$, is then dominated at low temperature by 
$s_D(T)\simeq (\pi/6)(T/J)$, while the domain wall contribution takes
over at higher temperature. To be more quantitative, let us define the
temperature $T_1$ where both contributions are equal. It is given by
$(E_{DW}/T_1)\exp(-E_{DW}/T_1)=(\pi/6)(T_1/J)$, which leads to $T_1=0.074 J$. 
This is
in very good agreement with the numerical results of Fig.~\ref{entropy}. 

The prediction for the density of domain walls is also in very good agreement
with the numerical simulations (see inset of Fig.~\ref{phase_nr}, dashed line) 
for not too low low temperatures:
A fit with an exponential law 
$p=\exp(-E_0/T)$ for $8\le\beta\le 12$ 
gives $E_0=0.36(1)$, in very good agreement with the domain-wall energy 
$E_{DW}$. 
For very low temperatures, namely for temperatures where the average number of
domain walls is of order 1 or smaller, and for finite systems, 
the above picture cannot work
because the numerical simulations were performed using periodic boundary
conditions, and domain walls can be created only by pairs with a minimum energy
$2E_{DW}$. Neglecting configurations with more than one pair of domain walls, 
one can show that the concentration is 
expected to behave like $p\simeq (L-1) \exp (-2E_{DW}/T)$. A fit of 
the very low temperature numerical data could indeed be performed with the law 
$p= (L-1) \exp (-E_1/T)$ for $15\le \beta \le 25$ (see inset of Fig.~\ref{phase_nr})  
with $E_1=0.72(2)$, again in very good
agreement with $2E_{DW}$. Note that the temperature $T_0$ below which
finite-size effects due to periodic boundary
conditions start to influence the thermodynamics is given by
$\exp(-E_{DW}/T_0)=1/L$, i.e. $T_0\simeq E_{DW}/\ln L$. Since it vanishes when 
the system size goes to infinity, the free energy of Eq. (\ref{free}) is expected to be
valid down to zero temperature in the thermodynamic limit.

To summarize this section, the very low temperature excitations correspond to
simultaneous flips of spin and orbital degrees of freedom within one domain, and
domain wall excitations corresponding to collective excitations involving spins
and orbital degrees of freedom play an important role above a cross-over
temperature $T^*\simeq 0.05 J$.

{\bf The dimerized XXZ model:} It corresponds to the  Hamiltonian of
Eq. (\ref{ham2}) with 
$J_i=J$ if $i$ is even and $J_i=\alpha J$ if $i$ is odd. We wish to study
that model in the limit $\alpha \ll 1$. Let us start by introducing some
notations. The Hilbert space of a given bond is spanned by the 16 states 
$|SS\rangle$,
$|ST_i\rangle$ ($i=-1,0,1$), $|T_iS\rangle$ ($i=-1,0,1$) and $|T_iT_j\rangle$ ($i,j=-1,0,1$),
where the first (second) letter refers to the spins 
(orbitals), while $|S\rangle$ and $|T_i\rangle$ are the usual singlet and triplets given by
$|S\rangle=(|\uparrow \downarrow\rangle - | \downarrow \uparrow\rangle)/\sqrt{2}$,
$|T_1\rangle=|\uparrow \uparrow \rangle$, 
$|T_0\rangle=(|\uparrow \downarrow\rangle + | \downarrow \uparrow\rangle)/\sqrt{2}$ and
$|T_{-1}\rangle=|\downarrow \downarrow\rangle$. In the pure Heisenberg case, the
six states $|ST_i\rangle$ and $|T_iS\rangle$ ($i=-1,0,1$) are degenerate groundstates.
However this degeneracy is partially lifted if $\lambda<1$ in Eq.(\ref{ham2}), 
and the groundstate of a given bond is only two--fold degenerate ($|ST_0\rangle$ and 
$|T_0S\rangle$). If $\alpha=0$, the groundstate of Eq.(\ref{ham2}) is then 
$2^{L/2}$--fold degenerate, where $L$ is the number of sites, since each dimer
$(i,i+1)$, $i$ even, can be in any of the two states $|ST_0\rangle_i$ or 
$|T_0S\rangle_i$. Let us study how this
degeneracy is lifted when $\alpha$ is switched on. Since we have a two-level
system on each dimer $(i,i+1)$, $i$ even, we can define a pseudo spin-1/2
operator $\vec \sigma_i$ that acts on this dimer with the identification
$|ST_0\rangle\equiv |\downarrow\rangle$ and $|T_0S\rangle\equiv |\uparrow\rangle$. An effective 
Hamiltonian can then be derived using standard many-body perturbation theory.
The result depends on $\lambda$. If $\lambda>0$, then the perturbation is lifted
to first order in $\alpha$, and the effective Hamiltonian reads
\begin{equation}
H_{eff}^{\lambda>0}=\alpha \lambda^2 \sum_{i\ {\rm even}} (\sigma_i^x 
\sigma_{i+2}^x + \frac{1} {4} ) 
\end{equation}
while if $\lambda=0$ one has to go to second order perturbation theory to lift
the degeneracy, and the effective Hamiltonian reads:
\begin{equation}
H_{eff}^{\lambda=0}= - \alpha^2 \sum_{i\ {\rm even}} (\sigma_i^x 
\sigma_{i+2}^x + \frac{1} {4} ) 
\end{equation}
Several remarks can be made about these results: First of all, the effective 
Hamiltonian is always an Ising model to the first non--vanishing order in
perturbation theory. Second, the effective Ising model is 
antiferromagnetic if 
$\lambda>0$ and ferromagnetic if $\lambda=0$. These models are of course 
related
by a simple transformation, but we expect to have a transition
line between these cases in the ($\alpha, \lambda$)--plane along which the
effective Hamiltonian presumably takes a more complicated form. 
Finally, and
more importantly, we obtain an Ising model in terms of $\sigma_i^x =
(\sigma_i^+ + \sigma_i^-)/2$, {\it not} $\sigma_i^z$. So the eigenstates must be
written in terms of the eigenstates of $\sigma_i^x$, namely 
$(|ST_0\rangle+|T_0S\rangle)/\sqrt{2}$ and $(|ST_0\rangle-|T_0S\rangle)/\sqrt{2}$. 
They are thus of the general form 
$2^{-L/4} \prod_{i\ {\rm even}} (|ST_0\rangle_i\pm |T_0S\rangle_i)$.

Let us now briefly discuss the low-energy properties. Quite generally, 
we expect to have a two-fold degenerate groundstate, and
a gapped excitation spectrum. More specifically,
the groundstates are given by
\begin{equation}
|GS\pm\rangle= 2^{-L/4} \prod_{i\ {\rm even}} (|ST_0\rangle_i\pm |T_0S\rangle_i)
\end{equation}
in the ferromagnetic case, and by
\begin{equation}
|GS\pm\rangle= 2^{-L/4} \prod_{i\ {\rm even}} (|ST_0\rangle_i\pm (-1)^{i/2}|T_0S\rangle_i)
\end{equation}
in the antiferromagnetic case. The first
excitations are obtained by replacing $(|ST_0\rangle-|T_0S\rangle)/\sqrt{2}$ (resp. 
$(|ST_0\rangle+|T_0S\rangle)/\sqrt{2}$) by $(|ST_0\rangle+|T_0S\rangle)/\sqrt{2}$ (resp. 
$(|ST_0\rangle-|T_0S\rangle)/\sqrt{2}$) in one of these groundstates with energy
$\alpha^2/2$ in the ferromagnetic case and $\alpha \lambda^2 /2$ in the
antiferromagnetic case. So there is indeed a gap in the spectrum. More
importantly, it is clearly impossible to separate spin and orbital degrees of
freedom since $(|ST_0\rangle+|T_0S\rangle)/\sqrt{2}$ and $(|ST_0\rangle-|T_0S\rangle)/\sqrt{2}$ are not
eigenstates of $(\vec S_i+\vec S_{i+1})^2$ or of 
$(\vec \tau_i + \vec \tau_{i+1})^2$, and the excitations are neither spin 
excitations nor orbital excitations. 
They are transitions between resonating valence-bond states that intimately mix
spin and orbital degrees of freedom.

It is also interesting to note that the correlation functions on a strong bond
($i$ even) 
$\langle\vec S_i.\vec S_{i+1}\rangle$, $\langle\vec \tau_i.\vec \tau_{i+1}\rangle$ and 
$\langle(\vec S_i.\vec S_{i+1})(\vec \tau_i.\vec \tau_{i+1})\rangle$ are all 
negative, as in
the
SU(4) symmetric case, which excludes
mean-field theory as a good starting point for the same reasons (see
Ref.\cite{frischmuth}).

{\bf Discussion:} Coming back to the original problem of the nature of
the excitations in the SU(4) symmetric model, let us put our results
in perspective.  In both cases studied above, exact results have been
obtained, and the low--energy excitations are neither spin nor orbital
excitations, but involve both spin and orbital degrees of freedom on
an equal footing. This is a clear indication of the breakdown
of mean-field theory. It strongly suggests that the model of Eq. (\ref{ham1})
also possesses such low--lying excitations. In particular, the
operators $\alpha_i^\pm=S_i^\pm\tau_i^\pm$ and
$\beta_i^\pm=S_i^\pm\tau_i^\mp$ of the XY case are linear combinations
of the operators $S^\alpha\tau^\beta$ which can be seen as generators
of the SU(4) algebra (see Ref.\cite{zhang}), and it is likely that at
least part of the low-lying modes of the SU(4) symmetric model will be
predominantly built on these generators and will retain the mixed
character observed here. Besides, the fact that the correlation functions
$\langle\vec S_i.\vec S_{i+1}\rangle$, $\langle\vec \tau_i.\vec \tau_{i+1}\rangle$ and 
$\langle(\vec S_i.\vec S_{i+1})(\vec \tau_i.\vec \tau_{i+1})\rangle$ are all 
negative on a given bond appears in the
XXZ case as a direct consequence of the local degeneracy between the
states (spin singlet $\times$ orbital triplet) and (spin triplet
$\times$ orbital singlet). So the picture that emerges is that the
symmetry between spin and orbital degrees of freedom has dramatic
consequences on the low--lying excitations: The system is not able to
choose between spin or orbital singlets or triplets, and the
excitations are an intricate mixture of spin and orbital degrees of
freedom.

To complete the picture, it will be useful to study the XXZ model in
the whole parameter range $0\le \alpha,\lambda \le 1$. The main issues
are: i) The evolution of the spectrum along the line
$(\alpha=1,\lambda=0)\rightarrow(\alpha=1,\lambda=1)$ joining the XY
case and the model of Eq. (\ref{ham1}); ii) The number of low--lying modes, and in
particular the presence of a gap as a function of $\alpha$ and $\lambda$;
iii) The nature of the effective model in the limit $\lambda=1$,
$\alpha\ll1$. Work is in progress along these lines.

We acknowldege very useful discussions with T. M. Rice and F.-C. Zhang.
One of us (B.F.) is also grateful for
financial support from the Swiss Nationalfonds. The calculations were
performed on the Intel Paragon at the ETH Z\"urich.

\end{document}